\documentclass[pdftex,a4paper,12pt]{article}
\usepackage{amssymb,amsmath,amsfonts,ntheorem,nomencl,mathrsfs}
\usepackage[arrow, matrix, curve]{xy}
\usepackage[latin1]{inputenc}
\newcommand{\IC}{\mathbb{C}}
\newcommand{\IR}{\mathbb{R}}

\newcommand{\ILL}{\mathscr{L}}

\newcommand{\IDD}{\mathscr{D}}

\newcommand{\IPP}{\mathscr{P}}

\newcommand{\IN}{\mathbb{N}}

\newcommand{\Id}{{\rm d}}

\newcommand{\f}{\frac}
\newcommand{\nn}{\nonumber}

\newtheorem{theorem}{THEOREM}[section]

\newtheorem{Lemma}[theorem]{Lemma}

\newtheorem{Corollary}[theorem]{Corollary}

\newtheorem{Remark}[theorem]{Remark}

\newtheorem{Theorem}[theorem]{Theorem}

\newtheorem{Proposition}[theorem]{Proposition}

\newtheorem{Definition}[theorem]{Definition}

\begin{document}
\begin{titlepage}
\title{Nonrelativistic hydrogen type stability problems on nonparabolic $3$-manifolds}
  \author{Batu Güneysu\footnote{E-Mail: gueneysu@math.hu-berlin.de}\\
   Institut für Mathematik\\
   Humboldt-Universität zu Berlin\\
   }
\end{titlepage}

\maketitle 

\begin{abstract} Let $M$ be a noncompact Riemannian $3$-manifold, satisfying some assumptions that guarantee the existence of a minimal positive Green's function $G:M\times M\to (0,\infty]$. We prove the following two stability results. Firstly, we show that there is a $C>0$ such that for all $\kappa\geq 0$, all generalized Laplacians $\IPP$ on $M$ with Lichnerowicz potential term\footnote{Of course, a generalized Laplacian $\IPP$ with $V_{\IPP}=0$ is nothing but a Bochner Laplacian with respect to a Hermitian vector bundle over $M$.} $V_{\IPP}=0$, and all $y\in M$ one has 
\begin{align}
\IPP -\kappa G(\bullet,y)\geq -C\kappa^2. \nn
\end{align}
Secondly, we prove that there are $C,\kappa_0 >0$ such that for all generalized Laplacians $\IPP$ on $M$ with $\IPP\geq 0$, $|V_{\IPP}|\in\mathrm{L}^2(M)$ and all $\Lambda>0$, $0\leq \kappa\leq \kappa_0\Lambda^2$, $y\in M$ one has
\begin{align}
\IPP -\kappa G(\bullet,y)+\Lambda \int_M\left\|V_{\IPP}(x)\right\|^2_x\mathrm{vol}(\Id x) \geq -C\kappa^2. \nn
\end{align}
The first inequality corresponds to the nonrelativistic stability of Hydrogen type problems on $M$ with magnetic fields when spin is neglected, whereas the second inequality corresponds to a restricted nonrelativistic stability of Hydrogen type problems when spin is taken into account. 
\end{abstract}

\section{Introduction}\label{intro}

Let us first recall some classical facts concerning the stability of nonrelativistic Hamiltonians in the Euclidean $\IR^3$. The reader may find these and many more result on stability problems in the Euclidean setting in \cite{lieb} and the references therein. Let $\beta=\sum_j\beta_j\Id x^j$ be a smooth real-valued $1$-form on $\IR^3$ and let $\kappa\geq 0$, $y\in\IR^3$. Then there is a unique self-adjoint realization $H(\beta,y,\kappa)$ of $-\f{1}{2}\sum_j (\partial_j+\mathrm{i}\beta_j)^2- \kappa G(\bullet,y)$ in the Hilbert space $\mathrm{L}^2(\IR^3)$, where $G(x,y)$ is the Coulomb potential. It is a classical fact that for some $C>0$, which does not depend on $\beta$ or $y$, one has
\begin{align}
H(\beta,y,\kappa)\geq -C \kappa^2.\label{zy0}
\end{align}
If $Z=\kappa/{(+e)}^2\in\IN$, then, up to certain conventions concerning the units and neglecting the electron's spin, $H(\beta,y,\kappa)$ is the nonrelativistic Hamiltonian of a Hydrogen type atom\footnote{Here and in the following, a Hydrogen type atom is understood to be an atom with one electron.} with $Z=\kappa/(+e)^2$ protons, in the magnetic field $\Id\beta$, where the nucleus is considered as fixed in $y$ with respect to the electron. In particular, the uniformity of the right hand side of (\ref{zy0}) in $\beta$ and $y$ has a simple but nevertheless important physical interpretation: Regardless of the number of protons, one cannot drive the energy of the atom to $-\infty$ by changing the magnetic field (or by simply moving the nucleus). Now if one takes the electron's spin nonrelativistically into account, the underlying differential expression has to be changed to the perturbed Pauli operator $\f{1}{2}\left(\sum_j\sigma^j (\partial_j+\mathrm{i}\beta_j)\mathbf{1}\right)^2-\kappa G(\bullet,y)\mathbf{1}$ in the Hilbert space $\mathrm{L}^2(\IR^3,\IC^2)$, with $\sigma^j$ the Pauli matrices. One natural way to get a unique self-adjoint realization $H_{\mathrm{Pauli}}(\beta,y,\kappa)$ of the latter differential operator is to assume that $\Id\beta$ has a finite self-energy (see for example proposition \ref{es} below):
\begin{align}
\int_{\IR^3}\left|\Id \beta \right|^2\Id x <\infty.\label{eay}
\end{align}
The electron's spin now has the effect that one only has a restricted stability: There is a  maximal $\kappa_0>0$ such that for all $0\leq \kappa\leq \kappa_0$ one has
\begin{align}
H_{\mathrm{Pauli}}(\beta,y,\kappa)+ \f{1}{8\pi}\int_{\IR^3}\left|\Id \beta \right|^2\Id x  \geq -C \kappa^2, \label{zy}
\end{align}
where $C>0$ does not depend on $\beta$ or $y$ (it is essential to shift the Hamiltonian here by the magnetic self-energy in order to get uniformity in $\beta$). This inequality has the same interpretation as before, with the difference that due to the maximality of $\kappa_0<\infty$, the atoms under consideration are only allowed to have a restricted number of protons. Otherwise, the atom can be made unstable by changing the magnetic field. \vspace{1.2mm}

The aim of this paper is to examine in detail which topological and Riemann geometric properties of the Euclidean $\IR^3$ actually guarentee the above stability results, in other words, we want to extend the above results to the setting of Riemannian $3$-manifolds. The essential observations in generalizing the above data to this setting are: \vspace{2mm}

(i) {\it There is a natural analogue of the Coulomb potential on any (necessarily noncompact) nonparabolic Riemannian manifold, namely, the minimal positive Green's function.} \vspace{1.2mm}

(ii) {\it Although, due to topological reasons, the concept of magnetic potentials does not make sense in general, one can nevertheless define magnetic Hamiltonians on arbitrary Riemannian manifolds.} \vspace{1.2mm}

(iii) {\it An appropriate use (see definition \ref{paul}) of the theory of Clifford modules makes it possible to generalize the Pauli Hamiltonian to $\mathrm{spin}^{\IC}$ manifolds.} \vspace{2mm}

Let us mention that (i) has already been observed in \cite{fro}, but we were particularly motivated by \cite{enciso}, where the author considers nonrelativistic many-body stability problems without magnetic fields on (necessarily noncompact) geodesically complete Riemannian $3$-manifolds with nonnegative Ricci curvature and an Euclidean volume growth. The observation (ii) is certainly well-known (see for example \cite{gruber} and \cite{laslo2}), and (iii) has been noticed in \cite{laslo} and \cite{fro}. \vspace{1mm}

Being equipped with (i)-(iii), we will prove natural extensions of (\ref{zy0}) and (\ref{zy}) to the class (A) of {\it noncompact geodesically complete Riemannian $3$-manifolds with a heat kernel that admits a global Gaussian upper bound.} This class is larger than the above mentioned class from \cite{enciso} (see remark \ref{bsp}). We believe it is a remarkable fact that this somewhat axiomatic assumption is the only explicit assumption that we have to make on the underlying Riemannian structure, in order to prove our main results. The point here is that the validity of a Gaussian upper bound has important analytic consequences like nonparabolicity, the validity of a Sobolev inequality or a maximal volume growth (see theorem \ref{nonp}).\vspace{1mm}

This work is organized as follows: After briefly recalling the concept of Green's functions in the setting of (A), we will prove a generalization of (\ref{zy0}) in theorem \ref{main} a), and a generalization of (\ref{zy}) in theorem \ref{main} b), both in the general setting of vector bundles over a Riemannian manifold of the type (A). The notion of generalized Laplacians (see definition \ref{zz}) will allow us to treat both cases in a symmetric way. A priori, theorem \ref{main} itself is a purely {\it mathematical} result, in the sense that the operators under consideration need not all correspond to physical situations. But we will show that theorem \ref{main} a) includes the nonrelativistic Hydrogen type magnetic Hamiltonians corresponding to ii) (corollary \ref{mag}), and that theorem \ref{main} b) includes the nonrelativistic Hydrogen type spin Hamiltonians corresponding to iii) (corollary \ref{spin}), so that we indeed arrive at the asserted {\it physical} generalizations of (\ref{zy0}) and (\ref{zy}) to the setting of Riemannian manifolds. Here, it should be noted that the constants $C,\kappa_0$ depend heavily on the underlying Riemannian structure and that the magnetic case is in complete analogy to (\ref{zy0}), whereas in the spin case, one has to add a ``generalized self-energy'' to the Hamiltonian, namely, the $\mathrm{L}^2$-norm square of the Lichnerowicz potential term (this term is a geometric generalization of the Zeeman term of the Euclidean Pauli operator above). Mathematically, this is no surprise and follows from the Lichnerowicz formula (lemma \ref{lich}), whereas on the physical side this fact may deserve some interpretation. In the above Euclidean situation, this ``generalized self-energy'' is nothing but the usual magnetic self-energy. \vspace{1mm}

Finally, let us remark that we have essentially considered one-body problems here, that is, atoms with one electron. It is certainly an interesting question to see to what extend our results, in particular in the spin case, can be extended appropriately to arbitrary molecules ('stability of matter'), in the setting of type (A) Riemannian manifolds.

\section{Main results}

Let $M$ be a smooth noncompact connected Riemannian $3$-manifold without boundary, where the Riemannian volume measure will be written as $\mathrm{vol}(\bullet)$ and the minimal positive heat kernel as $p_t(x,y)$.  For $x,y\in M$, $r>0$, the number $\Id(x,y)$ stands for the geodesic distance of $x,y$, and $\mathrm{K}_r(x)$ for the open geodesic ball with radius $r$ and center $x$. \\
If nothing else is said, $\mathrm{T}M$ and the corresponding tensor bundles are always considered as equipped with their standard Euclidean structure and the Levi-Civita connection $\nabla^{\mathrm{T}M}$. Furthermore, any data corresponding to $\mathrm{T}M$ will be implicitely complexified. \\
If $E\to M$ is a smooth (finite dimensional) Hermitian vector bundle, then with the usual abuse of notation the scalar product and norm corresponding to the Hilbert space $\Gamma_{\mathrm{L^2}}(M,E)$ will be written, respectively, as
\begin{align}
\left\langle f_1,f_2 \right\rangle =\int_M (f_1(x),f_2(x))_x \mathrm{vol}(\Id x),\>\>\left\|f\right\|:=\sqrt{ \left\langle f,f \right\rangle},\label{go}
\end{align}
where $(\bullet,\bullet)_x$ stands for the scalar product on $E_x$ and $\left\|\bullet\right\|_x=\sqrt{(\bullet,\bullet)_x}$ for the corresponding norm. The operator norms corresponding to these scalar products will also be denoted with $\left\|\bullet\right\|$ and with $\left\|\bullet\right\|_x$, respectively. It will also be convinient to introduce the notation 
\[
|\Psi|:M\longrightarrow  [0,\infty),\>\>|\Psi|(x):=\left\|\Psi(x)\right\|_x 
\]
for any section $\Psi$ in $E$ or in $\mathrm{End}(E)$. The symbol ${\dagger}$ stands for the formal adjoint of a differential operator between sections with respect to (\ref{go}), and finally, if $\nabla$ is a covariant derivative in $E$, then the corresponding curvature-$2$-form will be denoted with $\nabla^2\in \Omega^2(M,\mathrm{End}(E))$. \vspace{1.2mm}

We will first attack the problem of defining Coulomb type interactions in a general curved setting. To this end, we first recall the definition of Green's functions:

\begin{Definition} A measurable function $\tilde{G}:M\times M\to [-\infty,\infty]$ satisfying $\tilde{G}(x,y)=\tilde{G}(y,x)$, $\tilde{G}(x,\bullet)\in\mathrm{L}^1_{\mathrm{loc}}(M)$ and 
\begin{align}
-\Delta \tilde{G}(x,\bullet) = \delta_x \>\>\text{ for all $x,y\in M$}\label{edc}
\end{align}
is called a {\it Green's function} for $M$. Here, $\Delta=-\Id^{\dagger}\Id$ stands for the Laplace-Beltrami operator. 
\end{Definition}

In view of (\ref{edc}), the concept of Green's function only makes sense on noncompact Riemannian manifolds. Moreover, Green's functions always exist \cite{li} on noncompact Riemannian manifolds, however, they need not be unique. This is one of the motivations for the following definition: 

\begin{Definition} $M$ is called {\it nonparabolic}, if there exists a positive Green's function on $M$. 
\end{Definition}

The essential observation is (\cite{buch}, theorem 13.17):

\begin{Theorem} $M$ is nonparabolic, if and only if    
\[
\int^{\infty}_0 p_t(x,y) \Id t <\infty\>\>\text{ for some/all $x\ne y$.} 
\]
If this is the case, then 
\[
G:M\times M\longrightarrow  (0,\infty],\>\>G(x,y):= \int^{\infty}_0 p_t(x,y) \Id t
\]
is the (unique) minimal positive Green's function on $M$.
\end{Theorem}

Clearly, the Euclidean $\IR^3$ is nonparabolic and the corresponding minimal positive Green's function is given by (a positive constant times) the Coulomb potential $(x,y)\mapsto 1/|x-y|_{\IR^3}$. More generally, one has: 

\begin{Theorem}\label{nonp} Let $M$ be geodesically complete and assume that there are $C_1$, $C_2>0$ such that for all $t>0$, $x,y\in M$ one has
\begin{align}
p_t(x,y)\leq C_1t^{-\f{3}{2}}\mathrm{e}^{-\f{\Id(x,y)^2}{C_2 t}}.  \label{fd22}
\end{align}
Then the following assertions hold:\vspace{1.2mm}

{\rm a)} $M$ is nonparabolic and there is a $C_3>0$ with
\begin{align}
G(x,y)\leq \f{C_3}{\Id(x,y)} \text{ for all $x,y\in M$.}\label{marki0}
\end{align}
{\rm b)} There is a $C_4>0$ such that for any 
\begin{align}
h\in\mathrm{H}^{1,2}(M):=\left.\Big\{h\right|h\in\mathrm{L}^{2}(M), \nabla^{\mathrm{T}M} h\in \Gamma_{\mathrm{L}^{2}}(M, \mathrm{T}M)\Big\}\nn
\end{align}
one has 
 \begin{align}
\left(\int_M \left|h(x)\right|^{6}\mathrm{vol}(\Id x)\right)^{\f{1}{3}}\leq C_4\int_M\left\|\nabla^{\mathrm{T} M} h(x)\right\|^2_x\mathrm{vol}(\Id x).\label{sobo}
\end{align}
{\rm c)} One has an Euclidean volume growth, in the sense that
\begin{align}
\inf_{x\in M, r>0} \f{\mathrm{vol}(\mathrm{K}_r(x))}{r^3}>0.   \label{sjw}       
\end{align}
\end{Theorem}

{\it Proof.} a) We have 
\[
\int^{\infty}_0 p_t(x,y) \Id t\leq C_1 \int^{\infty}_0 t^{-\f{3}{2}}\mathrm{e}^{-\f{\Id(x,y)^2}{C_2 t}} \Id t = \f{4C_1\sqrt{\pi}}{\sqrt{C_2}\Id(x,y)}.
\]
b) This follows from the abstract fact that an estimate of the form $p_t(x,x)\leq C/ t^{3/2}$ always implies a Sobolev inequality of the form (\ref{sobo}) (see for example \cite{grig} and the references therein).\vspace{1.2mm}

c) Again, this is an abstract fact: The validity of (\ref{sjw}) is always implied by a Sobolev inequality of the form (\ref{sobo}) (\cite {saloff}, theorem 3.1.5). \vspace{0.5mm}

\hfill$\blacksquare$\vspace{2mm}

Being motivated from \cite{fro}, and in particular from \cite{enciso}, we will interpret $G(x,y)$ as the electrostatic potential that a point particle in $x\in M$ with charge $-e$ feels in the presence of a point particle in $y$ with charge $+e$. This again leads us to calling 
\[
G(\bullet,y):M\longrightarrow  (0, \infty]
\]
the {\it Coulomb potential} on $M$ with respect to the reference point $y\in M$.\\
In view of theorem \ref{nonp}, we will usually consider the following class of Riemannian $3$-manifolds : \vspace{3mm}

{\bf Assumption (A):} {\it $M$ is geodesically complete and there are $C_1$, $C_2>0$ such that for all $t>0$, $x,y\in M$ one has}
\begin{align}
p_t(x,y)\leq C_1 t^{-\f{3}{2}} \mathrm{e}^{-\f{\Id(x,y)^2}{C_2 t}}.\label{escc}
\end{align}

\begin{Remark}{\rm \label{bsp} If $M$ is geodesically complete with $\mathrm{Ric}\geq C$ and (\ref{sjw}), then (\ref{escc}) is satisfied if $\min\sigma(-\Delta)>0$ (this follows from a heat kernel estimate proven by E.B. Davies; see the last inequality on p. 31 in \cite{david}), or if one has $C=0$ (this is the usual Li-Yau heat kernel estimate; corollary 3.1 in \cite{liyau}). The Li-Yau case has been considered in the context of nonrelativistic many body systems without magnetic fields and spin in \cite{enciso}. 
}
\end{Remark}

We turn to the formulation of the generalized stability problem that we have in mind in this paper. In order to avoid any confusion, we add:

\begin{Definition}\label{zz} Let $E\to M$ be a smooth Hermitian vector bundle. A linear second order differential operator
\[
\IPP:\Gamma_{\mathrm{C}^{\infty}_0}(M,E)\longrightarrow \Gamma_{\mathrm{C}^{\infty}_0}(M,E)
\]
is called a generalized Laplacian on $M$, if $\IPP^{\dagger}=\IPP$ and if for any $x\in M$, $X\in \mathrm{T}_x M$, the operator symbol $\sigma_{\IPP,X}:E_x\to  E_x$ of $\IPP$ in $X$ satisfies 
\[
\sigma_{\IPP,X}(w)= -\left\|X\right\|^2_x w\>\>\text{ for all $w\in E_x$}.
\]
\end{Definition}

Let  
\[
\IPP:\Gamma_{\mathrm{C}^{\infty}_0}(M,E)\longrightarrow \Gamma_{\mathrm{C}^{\infty}_0}(M,E)
\]                                                   
be such a generalized Laplacian for the moment. A Lichnerowicz type theorem \cite{nico} implies the existence of a unique Hermitian covariant derivative $\nabla_{\IPP}$ in $E$ and a unique pointwise self-adjoint $V_{\IPP}\in \Gamma_{\mathrm{C}^{\infty}}(M,\mathrm{End}(E))$ such that 
\[
\IPP=\nabla^{\dagger}_{\IPP}\nabla_{\IPP} +V_{\IPP}: \Gamma_{\mathrm{C}^{\infty}_0}(M,E)\longrightarrow \Gamma_{\mathrm{C}^{\infty}_0}(M,E).
\]
In analogy to the Euclidean $\IR^3$, assumption (A) easily implies $G(\bullet,y)\in \mathrm{L}^2_{\mathrm{loc}}(M)$ (through theorem \ref{nonp} a)), so that for any real number $\kappa\geq 0$ and any $y\in M$, we can consider the differential operator
\begin{align}
\IPP-\kappa G(\bullet,y)\mathbf{1} =  \nabla^{\dagger}_{\IPP}\nabla_{\IPP} +V_{\IPP}- \kappa G(\bullet,y)\mathbf{1}\label{roh}
\end{align}
in $\Gamma_{\mathrm{L}^{2}}(M,E)$, which, defined initially on $\Gamma_{\mathrm{C}^{\infty}_0}(M,E)$, clearly gives a symmetric operator. For any appropriate measurable $b:\IR\to\IR$, the measurable self-adjoint section $b(V_{\IPP})$ in $\mathrm{End}(E)$ can be defined with the fiberwise spectral calculus of $E$, like for example $V^{\pm}_{\IPP}$. With these preparations, the following can be said directly about the differential operators given by (\ref{roh}) and their Schrödinger semigroups:

\begin{Proposition}\label{es} Assume (A) and $|V_{\IPP}|\in\mathrm{L}^2(M)$, that is,
\[
\int_M\left\|V_{\IPP}(x)\right\|^2_x\mathrm{vol}(\Id x)<\infty,
\]
and let $\kappa\geq 0$, $y\in M$. 

{\rm a)} The operator (\ref{roh}) is essentially self-adjoint and its closure $H(\IPP,\kappa,y)$ is semibounded from below. Furthermore, the quadratic form $q_{\IPP,\kappa,y}$ corresponding to $H(\IPP,\kappa,y)$ is given by 
\begin{align}
&\mathrm{D}\left(q_{\IPP,\kappa,y}\right)= \left\{f\left|f,\sqrt{V^+_{\IPP}}f\in\Gamma_{\mathrm{L}^{2}}(M,E),\nabla_{\IPP} f\in \Gamma_{\mathrm{L}^{2}}(M,E\otimes \mathrm{T}^*M) \right\}\right. , \nn\\
& q_{\IPP,\kappa,y}(f)=\int_M \left\|\nabla_{\IPP} f(x)\right\|^2_x\mathrm{vol}(\Id x)\nn\\
&\>\>\>\>\>\>\>\>\>\>+\int_M (V_{\IPP}(x)f(x),f(x))_x \mathrm{vol}(\Id x)-\kappa \int_M G(x,y)\left\|f(x)\right\|^2_x \mathrm{vol}(\Id x). \nn
\end{align}
{\rm b)} For any $t>0$, $2\leq p\leq\infty$ one has 
\begin{align}
 \mathrm{e}^{-t H(\IPP,\kappa,y)}\in\ILL\left(\Gamma_{\mathrm{L}^2}(M,E),\Gamma_{\mathrm{L}^p}(M,E)\right).\label{smoo}
\end{align}
In particular, if $f$ is an eigensection of $H(\IPP,\kappa,y)$, then $f\in\Gamma_{\mathrm{L}^p}(M,E)$ for all $2\leq p\leq\infty$. 
\end{Proposition}

{\it Proof.} a) The essential self-adjointness and semiboundedness from below will follow from combining theorem 2.13, remark 2.2 and theorem 2.3 in \cite{Br}, if we can show that (\ref{roh}) is semibounded from below on $\Gamma_{\mathrm{C}^{\infty}_0}(M,E)$. To this end, let the potential $W:M\to\mathrm{End}(E)$ be given by 
\[
W:=V_{\IPP}- \kappa G(\bullet,y)\mathbf{1}=V^+_{\IPP}- \left(V^-_{\IPP}+\kappa G(\bullet,y)\mathbf{1}\right)=:W_1-W_2. 
\]
Then using (\ref{marki0}) one easily finds 
\[
|W_2|\leq |V_{\IPP}|+ |\kappa G(\bullet,y)| \in \mathrm{L}^2(M)+ \mathrm{L}^{\infty}(M).
\]
But (\ref{fd22}) combined with proposition 2.8 in \cite{gue2} implies 
\[
 \mathrm{L}^2(M)+ \mathrm{L}^{\infty}(M)\subset \mathcal{K}(M),
\]
the Kato class of the Riemannian manifold $M$, so that we can use theorem 2.13 in \cite{gue2} to deduce that (\ref{roh}) is semibounded from below. The asserted form of the quadratic form now is also implied by the latter theorem. \vspace{1mm}

b) This follows from the proof of a) and theorem 2.16 b) in \cite{G2}.\vspace{0.5mm}

\hfill$\blacksquare$\vspace{2mm}

Let us also remark that the Schrödinger semigroups given by Hamiltonians as in proposition \ref{es} are given by Feynman-Kac type path integral formulae (theorem 2.11 in \cite{G2}). Indeed, (\ref{smoo}) has been proven by such a formula.\vspace{2mm}

We can now formulate the main result of this paper: 

\begin{Theorem}\label{main} The following assertions hold under (A): \vspace{1.2mm}

{\rm a)} There is a constant $C>0$ such that for all generalized Laplacians $\IPP$ on $M$ with $V_{\IPP}=0$ and all $\kappa\geq 0$, $y\in M$ one has  
\begin{align}
H(\IPP,\kappa,y)\geq - C\kappa^2.  \label{haupt0}
\end{align}

{\rm b)} There are constants $C,\kappa_0 >0$ such that for all generalized Laplacians $\IPP$ on $M$ with 
\[
\IPP\geq 0,\>\>\int_M\left\|V_{\IPP}(x)\right\|^2_x\mathrm{vol}(\Id x)<\infty,
\]
and all $\Lambda>0$, $0\leq \kappa\leq \kappa_0\Lambda^2$,  $y\in M$ one has  
\begin{align}
H(\IPP,\kappa,y)+ \Lambda \int_M\left\|V_{\IPP}(x)\right\|^2_x\mathrm{vol}(\Id x) \geq  -C\kappa^2. \label{haupt}
\end{align}
\end{Theorem}

Note that, as it should be clear from the formulation, we have {\it not} fixed a vector bundle in theorem \ref{main}, that is, the underlying vector bundles are allowed to varry.\\
The proof of theorem \ref{main} will be given in section \ref{bewe}. At this point, we only add some remarks on this proof, which is an extension of the methods from the proof of theorem 9.1 in \cite{lieb} to our geometric situation.

\begin{Remark}\label{skiz} {\rm 1. As a first step, we will prove a stability result of the form 
\begin{align}
H(-\Delta,\kappa,y)\geq-C \kappa^2\>\>\text{ in $\mathrm{L}^2(M)$}\label{gai0} 
\end{align}
in proposition \ref{was}. Note that even though the operators $H(-\Delta,\kappa,y)$ are bounded from below by proposition \ref{es}, an inequality of the form 
\[
H(-\Delta,\kappa,y)\geq -C(\kappa) 
\]
is not clear at all in the general manifold setting, due to the lack of unitary transformations in the Hilbert spaces under consideration. Furthermore, the proof of (\ref{haupt}) uses the asserted {\it quadratic} dependence of (\ref{gai0}) on $\kappa$. In the Euclidean space $\IR^3$, it follows from a space scaling argument that the ground state energy of $H(-\Delta,\kappa,y)$ is actually {\it equal} to a negative constant times $\kappa^2$.\\
Inequality (\ref{gai0}) will be derived from the fact that $G(\bullet,y)$ is in the underlying Kato class, namely, we will combine an inequality from the theory of Kato perturbations of Dirichlet forms (theorem 3.1 in \cite{peter}) with (\ref{fd22}) and the KLMN theorem. The reason for this procedure is that our assumptions on the Riemannian structure do not lead to a direct control over the volume form, so that one cannot apply classical $\mathrm{L}^p$-techniques diectly. We also remark that proposition \ref{was} has a natural generalization to regular Dirichlet forms. \vspace{1.2mm}

2. The only reason why one can expect some uniformity of (\ref{haupt0}) and (\ref{haupt}) in $\IPP$ at all is a quadratic form version of the Kato inequality (see proposition \ref{dia}). This fact combined with (\ref{gai0}) is essentially enough to derive (\ref{haupt0}). \vspace{1.2mm}

3. The proof of (\ref{haupt}) makes a direct use of the Sobolev inequality (\ref{sobo}). In fact, the key to the proof of (\ref{haupt}) is a combination of (\ref{sobo}) and (\ref{gai0}). \vspace{1.2mm}

4. Of course, the constants $C$ and $\kappa_0$ depend very sensitively on the Riemannian structure of $M$. We will discuss this dependence and the importance of the number $\Lambda$ in $(\ref{haupt})$ from the physical point of view after remark \ref{h} below.} 
\end{Remark}

We now come to the quantum mechanical interpretation of theorem \ref{main}. To this end, we will define the classes of magnetic Hamiltonians and Pauli Hamiltonians, which both correspond to subclasses of the class of generalized Laplacians, and apply theorem \ref{main} to these subclasses. \vspace{1.2mm}

The first definition is standard and corresponds to magnetic Hamiltonians: 

\begin{Definition} {\rm a)} A differential form $\beta\in\Omega^2(M)$ is called a magnetic field on $M$, if $\beta$ is real-valued and closed.\vspace{1.2mm}

{\rm b)} If $E\to M$ is a smooth Hermitian line bundle, then a generalized Laplacian 
\[
\IPP:\Gamma_{\mathrm{C}^{\infty}_0}(M,E)\longrightarrow  \Gamma_{\mathrm{C}^{\infty}_0}(M,E) 
\]
with $V_{\IPP}=0$ is called a magnetic Laplacian on $M$.  
\end{Definition}

Let $\IPP$ be a magnetic Laplacian. Firstly, note that the essential fact for this notion is that $\nabla^2_{\IPP}/\mathrm{i}$ is a magnetic field. Secondly, analogously to section \ref{intro}, one can interpret $H(\IPP,\kappa,y)$ as the nonrelativistic Hamiltonian corresponding to an atom on $M$ which is under the influence of the magnetic field $\nabla^2_{\IPP}/\mathrm{i}\in\Omega^2(M)$, and which has one electron (neglecting spin) and a nucleus with $Z=\kappa/{(+e)}^2$ protons. The nucleus is considered as fixed in $y$ with respect to the electron. Theorem \ref{main} a) directly implies: 

\begin{Corollary}\label{mag} Under (A), there is a constant $C> 0$ such that for all magnetic Laplacians $\IPP$ on $M$ and all $\kappa\geq 0$, $y\in M$ one has  
\begin{align}
H(\IPP,\kappa,y)\geq - C\kappa^2.  \label{hauptmag}
\end{align}
 \end{Corollary}

This result allows the following interpretation:

\begin{Remark} {\rm The fact that the rhs of (\ref{hauptmag}) is uniform in $\IPP$ and $y$ means that the energy of the Hydrogen type systems under consideration cannot be driven to $-\infty$ by changing $\IPP$, which corresponds to changing the magnetic field, or by moving the nucleus, whereas the arbitraryness of $\kappa$ here means that these facts are true for any atom, regardless of the number of protons. } 
\end{Remark}

We now come to the nonrelativistic description of a Hydrogen type atom, taking the electron's spin into account. Assume that $E\to M$ is a smooth Hermitian vector bundle with a {\it Clifford multiplication}
\[
c: \mathrm{T}^*M\longrightarrow\mathrm{End}(E),
\]
that is, $c$ is a morphism of smooth vector bundles such that for all $\alpha\in \Omega^1(M)$ one has
\begin{align}
c(\alpha)=-c(\alpha)^*,\>\>c(\alpha)^*c(\alpha)=\left|\alpha\right|^2.\label{gh0} 
\end{align}
Assume furthermore that $\nabla$ is a {\it Clifford connection} with respect to $c$, that is, $\nabla$ is a Hermitian covariant derivative in $E$ such that for all $\alpha$ as above and all $X\in \Gamma_{\mathrm{C}^{\infty}}(M,\mathrm{T}M),\psi\in\Gamma_{\mathrm{C}^{\infty}}(M,E)$ one has
\[
 \nabla_X(c(\alpha)\psi)=c(\nabla^{\mathrm{T} M}_X \alpha )\psi+ c(\alpha)\nabla_X \psi. 
\]
Then $(c,\nabla)$ is called a {\it Dirac structure} on $M$, and one can define the associated {\it Dirac operator} $\IDD(c,\nabla)$ by
\[
\IDD(c,\nabla):=c\circ \nabla:\Gamma_{\mathrm{C}^{\infty}_0}(M,E)\longrightarrow \Gamma_{\mathrm{C}^{\infty}_0}(M,E),
\]
which is a linear first order differential operator with $\IDD(c,\nabla)^{\dagger}=\IDD(c,\nabla)$. If $(e_j)$ is some smooth local orthonormal frame for $\mathrm{T}M$, then one has $\IDD(c,\nabla)=\sum_jc(e_j^*) \nabla_{e_j}$. Furthermore, $\IDD(c,\nabla)^2$ is a generalized Laplacian on $M$ which is given by a Lichnerowicz formula \cite{lawson}
\begin{align}
\IDD(c,\nabla)^2=\nabla^{\dagger}\nabla + \sum_{i<j} c(e^*_i)c(e^*_j) \nabla^2(e_i,e_j).\label{sgq}
\end{align}
In particular, $\IDD(c,\nabla)^2$ is a nonnegative generalized Laplacian. Being motivated from \cite{fro} and in particular from \cite{laslo}, we propose:

\begin{Definition}\label{paul} In the above situation, $(c,\nabla)$ is called a Pauli-Dirac structure on $M$, if $\mathrm{rank} E= 2$. In this situation, the operator $\IPP(c,\nabla):=\IDD(c,\nabla)^2$ is called the {\it Pauli operator} corresponding to $(c,\nabla)$.  
\end{Definition}

The assumption $\mathrm{rank}  E= 2$ is a topological restriction on $M$, namely, the existence of a Pauli-Dirac structure on $M$ is equivalent to $M$ being a $\mathrm{spin}^{\IC}$ manifold. This follows from $\dim M=3$ and $\mathrm{spin}^{\IC}(3)\cong\mathrm{U}(2)$ as Lie groups. Furthermore, one has: 

\begin{Lemma}\label{lich} Let $(c,\nabla)$ be a Pauli-Dirac structure on $M$. Then $\mathrm{tr} [\nabla^2]/ \mathrm{i}\in\Omega^2(M)$ is a magnetic field, and the Lichnerowicz formula (\ref{sgq}) takes the form
\begin{align}
\IPP(c,\nabla)= \nabla^{\dagger}\nabla + \f{1}{4}\mathrm{scal}(\bullet)\mathbf{1} +  \f{1}{2}\sum_{i<j} \mathrm{tr}\left[ \nabla^2\right](e_i,e_j)c(e^*_i)c(e^*_j).\label{ggs}
\end{align}
\end{Lemma}

{\it Proof.} These assertions are all included in proposition 11 and theorem 19 from \cite{laslo}. One just has to note that $\mathrm{i}c$ defines a $\mathrm{spin}^{\IC}$ spinor bundle, and that $\nabla$ determines a $\mathrm{spin}^{\IC}$ connection on $\mathrm{i}c$, both in the sense of \cite{laslo}. \vspace{0.5mm}

\hfill$\blacksquare$\vspace{2mm}

\begin{Remark}{\rm The fact that $\mathrm{tr} [\nabla^2]/\mathrm{i}$ is a magnetic field, and that only this scalar part of $\nabla$ appears in the potential term $V_{\IPP(c,\nabla)}$ are essential for the notion ''Pauli operator`` (see the differential expression for $H_{\mathrm{Pauli}}(...)$ from section \ref{intro}).  }
\end{Remark}

Assume now that the Riemannian structure of $M$ satisfies (A) and that $(c,\nabla)$ is a Pauli-Dirac structure on $M$ with 
\begin{align}
S(c,\nabla):=\int_M \left|\left\|\f{1}{4}\mathrm{scal}(\bullet)\mathbf{1} +  \f{1}{2}\sum_{i<j} \mathrm{tr}\left[ \nabla^2\right](e_i,e_j)c(e^*_i)c(e^*_j)\right\|\right|^2_x \mathrm{vol}(\Id x)<\infty,\nn
\end{align}
where $\left|\left\|\bullet \right\|\right|_x$ stands for the fiberwise Hilbert-Schmidt norm. Then for any $\kappa\geq 0$, $y\in M$, the operator 
\[
H(c,\nabla;\kappa,y):=H(\IPP(c,\nabla),\kappa,y) 
\]
is well-defined in the sense of proposition \ref{es}, which follows from the trivial inequality 
\begin{align}
\left|\left\|\bullet \right\|\right|_x\geq \left\|\bullet \right\|_x.\label{norm} 
\end{align}
In view of section \ref{intro}, this operator can be interpreted as the nonrelativistic Hamiltonian corresponding to a Hydrogen type atom on $M$ which has a nucleus with $Z=\kappa/{(+e)}^2$ protons and which is under the influence of the magnetic field $\mathrm{tr}[\nabla^2]/ \mathrm{i}$, where in contrast to the bare magnetic situation of corollary \ref{mag}, we have now taken spin into account. Again, the nucleus is considered as fixed in $y$ with respect to the electron. 

\begin{Corollary}\label{spin} Under (A), there are constants $C> 0$, $\kappa_0 >0$ such that for all Pauli-Dirac structures $(c,\nabla)$ on $M$ with $S(c,\nabla)<\infty$ and all $\Lambda>0$, $0\leq \kappa\leq \kappa_0\Lambda^2$,  $y\in M$ one has  
\begin{align}
&H(c,\nabla;\kappa,y)+ \Lambda S(c,\nabla)\geq  -C\kappa^2. \label{hauptspin}
\end{align}
\end{Corollary}

{\it Proof. } This follows directly from theorem \ref{main} b) and (\ref{norm}). \vspace{0.5mm}

\hfill$\blacksquare$\vspace{2mm}


This result can be interpreted as follows: 

\begin{Remark}\label{h} {\rm 1. That the rhs of (\ref{hauptspin}) does not depend on $(c,\nabla)$ or $y$ means that the energy of the system cannot be driven to $-\infty$ by changing the underlying Pauli-Dirac structure (which includes the magnetic field) or by moving the nucleus. \vspace{1.2mm}

2. In generalizing the Euclidean situation (\ref{zy}), we had to add a {\it generalized self-energy} $S(c,\nabla)$ to the Hamiltonian in order in order to obtain a stability result which only contains concrete geometric/physical data. It is easily checked that in the Euclidean situation that we have considered in section \ref{intro}, this generalized self-energy indeed reduces to nothing but the usual magnetic self-energy (\ref{eay}). \vspace{1.2mm}


3. The fact that in contrast to corollary \ref{mag}, (for a fixed value of $\Lambda$) the number $\kappa$ cannot be taken arbitrarily big now, comes from spin through the Lichnerowicz potential terms, and this numerical restriction physically means that a stability of the form (\ref{hauptspin}) is restricted to nuclei that have $Z\leq (\kappa_0\Lambda^2)/{(+e)}^2$ protons (see also section \ref{intro}). \vspace{1.2mm}

4. By choosing and fixing $\Lambda$ appropriately with respect to $\kappa_0$ once for all, one can always guarantee stability for $Z=1$, the classical Hydrogen case. }
\end{Remark}

Finally, note that in view of remark \ref{h}, it is clear that it is an important task to determine how big the constant $\kappa_0$, which depends very sensitively on the Riemannian structure of $M$, can be taken. Our choice of $\kappa_0$ is such that $\kappa_0$ becomes big, if the constant  $C_1$ from (\ref{fd22}) or the constant $C_4$ from (\ref{sobo}) becomes small. In other words: The question of finding good heat kernel estimates and good Sobolev estimates is closely related with quantum mechanical stability problems of the type (\ref{hauptspin}).

\section{Proof of theorem \ref{main}.}\label{bewe}

This section is devoted to the proof of theorem \ref{main}. In view of remark \ref{skiz}, we start with the following proposition, the simplest scalar form of theorem \ref{main}, which will be used several times in the proof of theorem \ref{main} (not only for uniformity in $y$!). We recall that 
\[
H(-\Delta,\kappa,y)=H(\Id^ {\dagger}\Id,\kappa,y)
\]
and $q_{-\Delta, \kappa,y}$ stand, respectively, for the self-adjoint realization and the corresponding quadratic form in the sense of proposition \ref{es} of $-\Delta-\kappa G(\bullet,y)$ in $\mathrm{L}^2(M)$.

\begin{Proposition}\label{was} Under (A), there is a $C_6> 0$ such that for all $\tilde{\kappa}\geq 0$, $y\in M$ and all $h\in\mathrm{H}^{1,2}(M)=\mathrm{D}(q_{-\Delta, \tilde{\kappa},y})$ one has 
\begin{align}
q_{-\Delta, \tilde{\kappa},y}(h)=q_{\Id}(h)-\tilde{\kappa} \int_M G(x,y)\left|h(x)\right|^2 \mathrm{vol}(\Id x) \geq -C_6 \tilde{\kappa}^2 \left\|h\right\|^2,
\end{align}
in other words, $H(-\Delta,\tilde{\kappa},y)\geq  -C_6 \tilde{\kappa}^2$. 
\end{Proposition}

{\it Proof.} We can assume $\tilde{\kappa}>0$ for the proof. For any $r>0$, $y\in M$ let
\[
C(r,y):=\sup_{x\in M}\int^{\infty}_0 \mathrm{e}^{-rs}\int_M p_s(x,z) G(z,y) \mathrm{vol}(\Id z) \Id  s\in [0,\infty].
\]
We will prove in a moment that there is a $C>0$ such that for all $r$,$y$ one has 
\begin{align}
C(r,y)\leq \f{C}{\sqrt{r}}.\label{qwq} 
\end{align}
In particular, $G(\bullet,y)$ is in the Kato class $\mathcal{K}(M)$ and the proof of theorem 2.13 in \cite{gue2} (it is here where theorem 3.1 in \cite{peter} is used heavily) shows that for all $r,\tilde{\kappa}$, $y$, $h$, 
\[
\tilde{\kappa} \int_M G(x,y)\left|h(x)\right|^2 \mathrm{vol}(\Id x) \leq \f{\tilde{\kappa}  C}{\sqrt{r}} q_{\Id}(h)+  \sqrt{r}\tilde{\kappa}  C\left\|h\right\|^2.
\]
Choosing for example $r=2C^2\tilde{\kappa}^2$ implies
\[
\tilde{\kappa} \int_M G(x,y)\left|h(x)\right|^2 \mathrm{vol}(\Id x) \leq \f{1}{\sqrt{2}} q_{\Id}(h)+  \sqrt{2}\tilde{\kappa}^2  C^2\left\|h\right\|^2, 
\]
so that the quadratic form defined by $\tilde{\kappa} G(\bullet, y)$ is $q_{\Id}$-form bounded with bound $<1$. In particular, the KLMN theorem X.17 in \cite{Re2} combined with the latter inequality implies
\[
H(-\Delta,\tilde{\kappa},y)\geq  - \sqrt{2} C^2 \tilde{\kappa}^2, 
\]
which proves the proposition. \\
It remains to prove (\ref{qwq}). The equality $G(z,y)= \int^{\infty}_0 p_t(z,y) \Id t$ combined with the Chapman-Kolmogorov identity for the heat kernel and (\ref{fd22}) give
\begin{align}
&\int^{\infty}_0 \mathrm{e}^{-rs}\int_M p_s(x,z) G(z,y) \mathrm{vol}(\Id z) \Id  s= \int^{\infty}_0 \int^{\infty}_0 \mathrm{e}^{-rs} p_{s+t}(x,y) \Id t \ \Id  s\nn\\
&\leq  C_1 \int^{\infty}_0 \mathrm{e}^{-rs}\left(\int^{\infty}_0 \f{1}{{(s+t)}^{\f{3}{2}}} \Id t \right) \Id  s= C_1 \int^{\infty}_0 \mathrm{e}^{-rs}\left(\int^{\infty}_s  \f{1}{t^{\f{3}{2}}} \Id t \right)\Id  s\nn\\
&\leq C'\int^{\infty}_0 \mathrm{e}^{-rs}  \f{1}{\sqrt{s}}  \Id  s= C''\int^{\infty}_0 \mathrm{e}^{-rs^2}   \Id  s  = \f{C}{\sqrt{r}},
\end{align}
and the proof is complete. \vspace{0.5mm}

\hfill$\blacksquare$\vspace{2mm}

Before we come to the proof of theorem \ref{main}, we state our second main tool: Kato's inequality (see \cite{gue2} for a proof in this formulation):

\begin{Proposition}\label{dia} Let $E\to M$ be a smooth Hermitian vector bundle and let $\nabla$ be a Hermitian covariant derivative in $E$. If $q_{\nabla}\geq 0$ denotes the quadratic form corresponding to the Friedrichs realization of $\nabla^{\dagger}\nabla$ in $\Gamma_{\mathrm{L}^2}(M,E)$ and if $q_{\Id}\geq 0$ denotes the quadratic form corresponding to the Friedrichs realization of  $-\Delta=\Id^{\dagger}\Id$ in $\mathrm{L}^2(M)$, then for any $f\in\mathrm{D}(q_{\nabla})$ one has $|f|\in\mathrm{D}(q_{\Id})$ and $q_{\Id}(|f|)\leq q_{\nabla}(f).$ 
\end{Proposition}

Note that proposition \ref{dia} is valid on any Riemannian manifold. Being equipped with the above results, we finally come to the proof theorem \ref{main}:\vspace{2mm}

{\it Proof of theorem \ref{main}.} We start with observations that will be relevant for both parts of the theorem: Let $\IPP$ be a generalized Laplacian on $M$ with $\IPP\geq 0$ and 
\begin{align}
\int_M\left\|V_{\IPP}(x)\right\|^2_x\mathrm{vol}(\Id x)<\infty,\label{endl}
\end{align}
and let $\kappa\geq 0$, $y\in M$, $\Lambda>0$, $f\in \mathrm{D}(q_{\IPP,\kappa,y})$. Then $\IPP\geq 0$ combined with 
\[
(V_{\IPP}(x)f(x),f(x))_x\geq -\left\|V_{\IPP}(x)\right\|_x \left\|f(x)\right\|^2_x 
\]
implies the following estimate for any $0\leq t \leq 1$,  
\begin{align}
&q_{\IPP,\kappa,y}(f) + \Lambda \int_M\left\|V_{\IPP}(x)\right\|^2_x\mathrm{vol}(\Id x)\nn\\
&= q_{\nabla_{\IPP}}(f)+\int_M (V_{\IPP}(x)f(x),f(x))_x \mathrm{vol}(\Id x) - \kappa \int_M G(x,y)\left\|f(x)\right\|^2_x \mathrm{vol}(\Id x)\nn\\
&\>\>\>\>+\Lambda \int_M\left\|V_{\IPP}(x)\right\|^2_x\mathrm{vol}(\Id x) \nn\\
&\geq  tq_{\nabla_{\IPP}}(f)-t\int_M \left\|V_{\IPP}(x)\right\|_x \left\|f(x)\right\|^2_x \mathrm{vol}(\Id x)\nn\\
&\>\>\>+\Lambda\int_M \left\|V_{\IPP}(x)\right\|^2_x\mathrm{vol}(\Id x)- \kappa \int_M G(x,y)\left\|f(x)\right\|^2_x \mathrm{vol}(\Id x).\label{un}
\end{align}
Applying proposition \ref{dia} to $\nabla=\nabla_{\IPP}$ gives $|f|\in \mathrm{D}(q_{\Id})=\mathrm{H}^{1,2}(M)$, and we can estimate (\ref{un}) by
\begin{align}
&\geq  tq_{\Id}(|f|)-t\int_M \left\|V_{\IPP}(x)\right\|_x \left\|f(x)\right\|^2_x \mathrm{vol}(\Id x)\nn\\
&+ \Lambda\int_M \left\|V_{\IPP}(x)\right\|^2_x\mathrm{vol}(\Id x)- \kappa \int_M G(x,y)\left\|f(x)\right\|^2_x \mathrm{vol}(\Id x). \label{allg}
\end{align}

a) With the above preparations, there is essentially nothing left to prove: It has to be shown that there is a $C>0$ such that for any $\kappa\geq 0$, $y\in M$, any generalized Laplacian $\IPP$ on $M$ with $V_{\IPP}=0$ and any $f\in\mathrm{D}(q_{\IPP,\kappa,y})$ with $\left\|f\right\|=1$ one has $q_{\IPP,\kappa,y}(f)  \geq -C\kappa^2$. But given such $\kappa$, $y$, $\IPP$, $f$, putting $t=1$ in (\ref{allg}) implies
\begin{align}
q_{\IPP,\kappa,y}(f) \geq q_{\Id}(|f|) - \kappa \int_M G(x,y)\left\|f(x)\right\|^2_x \mathrm{vol}(\Id x),
\end{align}
so that one can define $C$ to be the constant $C_6$ from proposition \ref{was}. \vspace{2mm}

b) It is sufficient to show that there are $C>0,\kappa_0 >0$ such that for any $\Lambda>0$, $0\leq \kappa\leq\kappa_0 \Lambda^2$, $y\in M$, any generalized Laplacian $\IPP$ on $M$ with $\IPP\geq 0$ and (\ref{endl}) and any $f\in\mathrm{D}(q_{\IPP,\kappa,y})$ with $\left\|f\right\|=1$ one has 
\begin{align}
q_{\IPP,\kappa,y}(f) +  \Lambda \int_M\left\|V_{\IPP}(x)\right\|^2_x\mathrm{vol}(\Id x)  \geq -C\kappa^2. \label{he}
\end{align}
Let $\Lambda$, $y$, $\IPP$, $f$ be given as above, let $\kappa \geq 0$ and $0\leq t\leq 1$. Since $|f|\in \mathrm{H}^{1,2}(M)$ as we have noted above, it follows from the Sobolev inequality (\ref{sobo}) and Cauchy-Schwarz that $|f|\in\mathrm{L}^6(M)\cap \mathrm{L}^4(M)$. Then, using $(1/\Lambda) \times$ the trivial inequality
\[
- t \Lambda \left\|V_{\IPP}(x)\right\|_x\left\|f(x)\right\|^2_x+  \Lambda^2\left\|V_{\IPP}(x)\right\|^2_x\geq -  t^2\left\|f(x)\right\|^4_x,
\]
we can estimate (\ref{allg}) by 
\begin{align}
\geq  tq_{\Id}(|f|)- \f{t^2}{\Lambda} \int_M  \left\|f(x)\right\|^4_x\mathrm{vol}(\Id x)- \kappa \int_M G(x,y)\left\|f(x)\right\|^2_x \mathrm{vol}(\Id x).\label{ghas}
\end{align}
As the function $\IR\to \IR$, $s\mapsto sq_{\Id}(|f|)- s^2\Lambda^{-1} \int \left|f\right|^4$ has its maximum in 
\[
s^*:=\f{ \Lambda q_{\Id}(|f|)}{2}\left( \int_M  \left\|f(x)\right\|^4_x\mathrm{vol}(\Id x)\right)^{-1},
\]
we proceed with the following case differentiation: If $s^*> 1$, then 
\[
 - \f{1}{\Lambda}\int_M  \left\|f(x)\right\|^4_x\mathrm{vol}(\Id x) \geq -\f{1}{2} q_{\Id}(|f|)
\]
and applying (\ref{ghas}) with $t=1$ and proposition \ref{was} with $\tilde{\kappa}=2\kappa$, $h=|f|$ implies
\begin{align}
q_{\IPP,\kappa,y}(f) + \Lambda \int_M\left\|V_{\IPP}(x)\right\|^2_x\mathrm{vol}(\Id x)\geq - 2C_6 \kappa^2.\label{uny} 
\end{align}
It remains to consider the case $s^*\leq  1$. Then one has
\[
\f{\Lambda }{2}q_{\Id}(|f|) \leq \int_M  \left\|f(x)\right\|^4_x\mathrm{vol}(\Id x) 
\]
and applying (\ref{ghas}) with $t=s^*$ becomes
\begin{align}
\geq \f{ \Lambda^2 q_{\Id}(|f|)^2}{4 \int_M  \left\|f(x)\right\|^4_x\mathrm{vol}(\Id x)}- \kappa \int_M G(x,y)\left\|f(x)\right\|^2_x \mathrm{vol}(\Id x).\label{qyp} 
\end{align} 
Next, combining Cauchy-Schwarz with $\left\|f\right\|=1$ and the Sobolev inequality (\ref{sobo}) implies
\begin{align}
\int_M \left\|f(x)\right\|^{4}\mathrm{vol}(\Id x) \leq C_4 \left(\int_M  \left\|\nabla^{\mathrm{T} M}|f|(x)\right\|^2_x\mathrm{vol}(\Id x)\right)^{\f{3}{2}}  =C_4 q_{\Id}(|f|)^{\f{3}{2}},
\end{align}
so that (\ref{qyp}) is
\begin{align}
\geq \f{ \Lambda^2\sqrt{q_{\Id}(|f|)}}{4 C_4}- \kappa \int_M G(x,y)\left\|f(x)\right\|^2_x \mathrm{vol}(\Id x). \label{ende}
\end{align}
Applying proposition \ref{was} with $h=|f|$ and 
\[
\tilde{\kappa}=D\int_M G(x,y)\left\|f(x)\right\|^2_x \mathrm{vol}(\Id x) 
\]
where $D>0$ is arbitrary, shows (using $\left\|f\right\|=1$ again) 
\begin{align}
q_{\Id}(|f|)\geq  (-C_6D^2+D)\left(\int_M G(x,y)\left\|f(x)\right\|^2_x \mathrm{vol}(\Id x)\right)^{2}.\label{gqy}
\end{align}
With the choice $D:=(2C_6)^{-1}$ one has $(-C_6D^2+D)=(4C_6)^{-1} >0$ and (\ref{gqy}) implies
\[
\sqrt{q_{\Id}(|f|)}\geq \f{1}{\sqrt{4C_6}}\int_M G(x,y)\left\|f(x)\right\|^2_x \mathrm{vol}(\Id x),
\]
so that combining the latter inequality with (\ref{ende}) we arrive at
\begin{align}
&q_{\IPP,\kappa,y}(f) + \Lambda \int_M\left\|V_{\IPP}(x)\right\|^2_x\mathrm{vol}(\Id x) \nn\\
 \geq   &\left(\f{   \Lambda^2    }{8\sqrt{C_6}C_4}-\kappa\right) \int_M G(x,y)\left\|f(x)\right\|^2_x \mathrm{vol}(\Id x),\nn
\end{align}
which is $\geq 0$, if 
\[
\kappa \leq \f{\Lambda^2}{8\sqrt{C_6}C_4}. 
\]
In view of the latter fact and inequality (\ref{uny}), the claim follows from taking $C:=  2C_6$ and 
\[
\kappa_0:=\f{\Lambda^2}{8\sqrt{C_6}C_4}. 
\]
\vspace{0.5mm}

\hfill$\blacksquare$\vspace{2mm}

{\it Acknowledgements} The author would like to thank Evgeny Korotyaev for very helpful discussions on Schrödinger operators. The research has been financially supported by the SFB 647: Raum - Zeit - Materie.


\begin{thebibliography}{99}


\bibitem{Br} Braverman, M. \&  Milatovich, O. \& Shubin, M.: {\it Essential self-adjointness of Schrödinger-type operators on manifolds.} Russian Math. Surveys  57  (2002),  no. 4, 641--692.


\bibitem{david} {Davies, E. B.:} {\it Gaussian upper bounds for the heat kernels of some second-order operators on Riemannian manifolds.} J. Funct. Anal.  80  (1988),  no. 1, 16--32.





\bibitem{enciso} Enciso, A.: {\it Coulomb Systems on Riemannian Manifolds and Stability of Matter.} Ann. Henri Poincare 12 (2011), 723--741.

\bibitem{laslo} Erdös, L. \& Solovej, J.P.: {\it The kernel of Dirac operators on $S^3$ and $R^3$.} Rev. Math. Phys. 13 (2001), no. 10, 1247--1280.

\bibitem{laslo2} Erdös, L.: {\it Recent developments in quantum mechanics with magnetic fields.} Spectral theory and mathematical physics: a Festschrift in honor of Barry Simon's 60th birthday, 401--428, Proc. Sympos. Pure Math., 76, Part 1, Amer. Math. Soc., Providence, RI, 2007.

\bibitem{fro} Fröhlich, J. \& Grandjean, O. \& Recknagel, A.: {\it Supersymmetric quantum theory, non-commutative geometry, and gravitation.} Symétries quantiques (Les Houches, 1995),  221--385, North-Holland, Amsterdam, 1998.


\bibitem{gruber} Gruber, M.: {\it Bloch theory and quantization of magnetic systems.} J. Geom. Phys.  34  (2000),  no. 2, 137--154.

\bibitem{buch} Grigor'yan, A.: {\it Heat kernel and analysis on manifolds.} AMS/IP Studies in Advanced Mathematics, 47. American Mathematical Society, Providence, RI; International Press, Boston, MA, 2009.

\bibitem{grig} Grigor'yan, A.: {\it Heat kernel on a non-compact Riemannian manifold.} Proceedings of Symposia in Pure Mathematics, 57 (1995), 239-263.


\bibitem{gue2} Güneysu, B.: {\it Kato's inequality and form boundedness of Kato potentials on arbitrary Riemannian manifolds.} Preprint.


\bibitem{G2} Güneysu, B.: {\it On generalized Schrödinger semigroups.} To appear in Journal of Functional Analysis. 







 



\bibitem{lawson} Lawson, H. B. Jr. \& Michelsohn, M.-L.: {\it Spin geometry.} Princeton Mathematical Series, 38. Princeton University Press, Princeton, NJ, 1989.


\bibitem{liyau} Li, P. \& Yau, S.T.: {\it On the parabolic kernel of the Schrödinger operator.} Acta Math.  156  (1986),  no. 3-4, 153--201.

\bibitem{li} Li, P. \& Tam, L.F.: {\it Symmetric Green's functions on complete manifolds.} Amer. J. Math. 109 (1987), 1129--1154.

\bibitem{lieb} Lieb, E. \& Seiringer, R.: {\it The stability of matter in quantum mechanics. } Cambridge University Press 2009.

\bibitem{nico} Nicolaescu, L.I.: {\it Lectures on the geometry of manifolds.} Second edition. World Scientific Publishing Co. Pte. Ltd., Hackensack, NJ, 2007.



\bibitem{Re2} Reed, M. \& Simon, B.: {\it Methods of modern mathematical physics. II. Fourier analysis, self-adjointness.} Academic Press, Inc., 1975.


\bibitem{saloff} Saloff-Coste, L.: {\it Aspects of Sobolev-type inequalities.} London Mathematical Society Lecture Note Series, 289. Cambridge University Press, Cambridge, 2002.

\bibitem{peter} Stollmann, P. \& Voigt, J.: {\it Perturbation of Dirichlet forms by measures.} Potential Anal. 5 (1996), no. 2, 109--138.



\end{thebibliography}
\end{document}